\documentclass[journal]{IEEEtran}
\ifCLASSINFOpdf
\else
\fi
\usepackage{amsmath}
\usepackage{graphicx}
\usepackage{hyperref}
\usepackage{booktabs}
\usepackage{multirow}
\usepackage{makecell}
\usepackage{cite}

\makeatletter
\renewcommand\subsubsection{\@startsection{subsubsection}{3}{\parindent}% Set the indentation to match paragraph indent
  {0.1\baselineskip} % Space before the subsubsection title
  {0.1\baselineskip} % Space after the subsubsection title
  {\normalfont\normalsize\itshape}} % Formatting for the subsubsection title
\makeatother

\begin{document}
\title{Stochastic Black Start Resource Allocation to Enable Dynamic Formation of Networked Microgrids and DER-aided Restoration}

\author{Cong~Bai,~\IEEEmembership{Student~Member,~IEEE,} Salish~Maharjan, Han~Wang,~\IEEEmembership{Member,~IEEE,} and~Zhaoyu~Wang,~\IEEEmembership{Senior~Member,~IEEE}% <-this % stops a space
% \thanks{This work was funded by the U.S. Department of Energy Solar Energy Technologies Office under Agreement Number 40385}% <-this % stops a space
}

\maketitle
\begin{abstract}
Extended outages in distributed systems (DSs) dominated by distributed energy resources (DERs) require innovative strategies to efficiently and securely deploy black start (BS) resources. To address the need, this paper proposes a two-stage stochastic resource allocation method within synchronizing dynamic microgrids (MGs) for black start (SDMG-BS), enabling risk-averse and adaptive restoration across various scenarios while ensuring frequency security. Virtual synchronous generator (VSG)-controlled grid-forming inverters (GFMIs) equipped with primary frequency governors (PFGs) are modeled as BS resources. Their frequency response is characterized by three transient indices, which are deployed as frequency dynamic constraints on load pick-up events to ensure frequency stability during the BS process. SDMG-BS framework facilitates location-independent synchronization among restored MGs and with the transmission grid (TG) with the help of smart switches (SSWs). The model incorporates scenario-based stochastic programming to address multi-source uncertainties, including season-dependent operational conditions and unpredictable TG outage durations, ensuring a resilient allocation plan. The proposed approach is validated on a modified IEEE 123-node feeder with three study cases designed across sixteen uncertainty scenarios.
\end{abstract}

\IEEEpeerreviewmaketitle
\section{Introduction}
\IEEEPARstart{F}{requent} transmission grid (TG) failures caused by severe weather and cyberattacks have led to an increasing number of blackouts in downstream distribution systems (DSs)~\cite{Avraam2023}. As a result, the self-start capability of DSs supported by the black start (BS) technique is essential for managing outages and enhancing the system’s resilience without relying on external resources. However, the traditionally used BS resources, such as diesel generators (DGs), are costly, necessitating a novel and efficient BS strategy for modern distributed energy resource (DER)-led DSs~\cite{Konar2023}.

BS resource allocation involves optimally placing and sizing resources that can restore the DS while ensuring security and efficiency~\cite{Zhang2019,Yao2020}. Unlike traditional methods that rely solely on DGs, DER-based BS strategies integrate diverse resources for rapid and comprehensive system recovery. Renewable energy sources (RESs), such as photovoltaic (PV) systems and wind turbines (WTs), are typically connected to the grid through grid-following inverters (GFLIs), which lack self-start capability~\cite{Zheng2024}. To energize these RESs and leverage them for restoration, DSs utilize battery energy storage systems (BESSs) with grid-forming inverters (GFMIs) to form microgrids (MGs) during the recovery process~\cite{Du2020}. The optimal location and sizing of GFMI-based BESSs are crucial for effective BS strategies. Additionally, advancements in communication and control technologies have led to the development of smart switches (SSWs) capable of dynamic reconfiguration, further enhancing the flexibility of DER-based BS strategies~\cite{Wu2023}.
A two-level BS framework is proposed in~\cite{HeidariAkhijahani2023}, where a 7-th-order transient simulation model for the GFMI-based DER is created to address frequency stability. However, this approach requires support from a dynamic simulation model to estimate the frequency nadir, which can be computationally complex for long-term planning problems like BS resource allocation. Additionally, it overlooks the rate of Change of Frequency (RoCoF). Similarly, a simulation-assisted service restoration model is presented in~\cite{Zhang2021}, where a droop-based controller for the GFMI-based DER is adopted to deal with the frequency fluctuations during the BS. To handle the restoration of a low inertia power system, a fast frequency response enabled bi-level optimization model is established in~\cite{Qin2024}, where the RoCoF and the frequency nadir limit the maximum imbalanced power of the system. This model does not put constraints on frequency directly while the quasi-steady frequency regulation of DER is neglected. To fully utilize the potential of GFMI-based DER and overcome significant frequency variations during the BS, a comprehensive dynamic model that includes multiple transient indices is currently lacking in the literature.

Synchronization among restored MGs and with the TG is another critical aspect of BS operations. Many existing works ~\cite{HeidariAkhijahani2023,Zhang2021,Qin2024,Ding2022} form isolated MGs at the end of the restoration process, but their continuous operation is questionable when relying solely on DER-based BS and restoration. A frequency-based synchronization is necessary for the continuous operation of MGs supported by DERs. A reactive power synchronization method for BS supported by PV systems is proposed in~\cite{PenaAsensio2023} to avoid the impact of active power variations on synchronism. However, this method restricts the optimization of synchronization during BS. To mitigate the inrush energization current and supply a smooth synchronization with neighboring MGs, a modified virtual synchronous generator (VSG) controller is designed for GFMI-based DERs in~\cite{Alassi2023}, validated in a real-time platform. Still, this method prohibits the assistance between the restored MGs and only relies on a fixed synchronization scheme after the entire system is energized. Pioneering works in~\cite{Du2022,Wang2024} have proposed efficient BS strategies for the DS, where the dynamic boundaries for networked MGs allow synchronization during the restoration. However, these studies depend on the location of MGs and the structure of DS, which restricts the flexibility of the choice of BS path. A more elastic synchronizable BS method is needed to find optimal cranking paths for BS at the resource allocation stage that considers the location and size of GFMIs as a decision variable.

Existing BS resource allocation models often employ deterministic approaches, failing to address uncertainties such as varying RES output, fluctuating load demands, and random fault events~\cite{Yao2020,Qiu2016,Patsakis2018}. A model predictive control-based restoration framework is proposed in~\cite{Zhao2018}, where the uncertainties of BS resources in the MG are handled with representative scenarios derived from discrete probability distributions of the forecast errors. Multi-source uncertainties, including the time-varying RES output, load demand, and random fault events, are considered in~\cite{Zou2020} to restore the DS and improve the system's reliability. In addition, the upstream TG outage duration is also a random event impacting the BS process. Ignoring these uncertainties can lead to suboptimal resource allocation decisions, which may prolong the DS restoration time. Incorporating these uncertainties is critical for robust and effective BS strategies.

Therefore, this paper proposes a frequency-constrained two-stage stochastic resource allocation model within synchronizable dynamic microgrids for black start (SDMG-BS) to address the multi-source uncertainties of BS resource allocation. This model considers the dynamic performance and coordination among restored MGs and with the TG. In summary, the technical contributions of this paper are as follows:\begin{itemize}
    \item A dynamic frequency response model for GFMI-based, VSG-embedded BESS with the primary frequency governor (PFG) is developed, introducing constraints on transient frequency indices to enhance MG resilience and frequency security during BS.
    \item A SDMG-BS framework is proposed to facilitate the restoration of DER-domained DSs. MGs supported by GFMI-based BESS dynamically synchronize at the same frequency through SSWs, accelerating and improving overall recovery.
    \item A two-stage stochastic resource allocation model in the SDMG-BS is fomulated to determine the optimal siting and sizing of GFMI-based BESSs and placement of SSWs, incorporating uncertainties such as RES output, load demand, and TG outage duration to ensure robust and adaptive restoration.
\end{itemize}

The remainder of this paper is organized as follows. Section~\ref{se:2} presents the modeling of critical equipment employed in DS restoration. Section~\ref{se:3} details the formulation of the proposed two-stage stochastic resource allocation model for SDMG-BS. And then, the simulation system and generated uncertainty scenarios are described in Section~\ref{se:4}. Further, three cases are designed and studied to validate the developed framework in Section~\ref{se:5}. Finally, Section~\ref{se:6} concludes the whole work.
\section{Modeling of Critical Equipment}\label{se:2}
In this section, firstly, the sets involved in the BS are defined. Secondly, the dynamic frequency response of the GFMI-based, VSG-controlled BESS with the PFG is modeled. Thirdly, the cold load pick-up (CLPU) effect of the restoration is described. Lastly, the SSW capable of synchronization is depicted.
% And then, the solution methodology for the proposed model is developed in Section~\ref{se:4}. Further, the simulation system parameters are presented, and three cases are designed and studied to validate the developed framework in Section~\ref{se:5}.
\vspace{-1em}
\subsection{Preliminary Sets}
For a given DS, denoted as $\Gamma = (\mathcal{B},\mathcal{L}, \Phi)$, where $\mathcal{B}$, $\mathcal{L}$, and $\Phi$ represent the sets of buses, branches, and phases, respectively, the DS is partitioned into multiple segments, referred as bus blocks. These segments, connected via switches, are represented by the set $\mathcal{G}$. 
Certain segments, identified by the subset $\mathcal{V}$, are selected as candidate locations for BESS installations, facilitating the formation of MGs due to their self-starting capability during BS. A set $\mathcal{W}$ is further defined to ensure that each synchronization occurs only between two distinct MGs, expressed as:
\begin{equation}\label{eq:twoelementinV}
    \mathcal{W} = \left\{\{k,l\} | k\in\mathcal{V}, l\in\mathcal{V}, k\neq l\right\}.
\end{equation}

The restoration process is modeled over an optimization horizon, represented by the set $\mathcal{T}$, while scenario-based uncertainties are described by the set $\mathcal{O}$.
\subsection{Modeling of GFMI-based BESS}
\subsubsection{VSG-controlled BESS with the PFG}
\begin{figure}[htbp]
\centering
\includegraphics[width=0.48\textwidth]{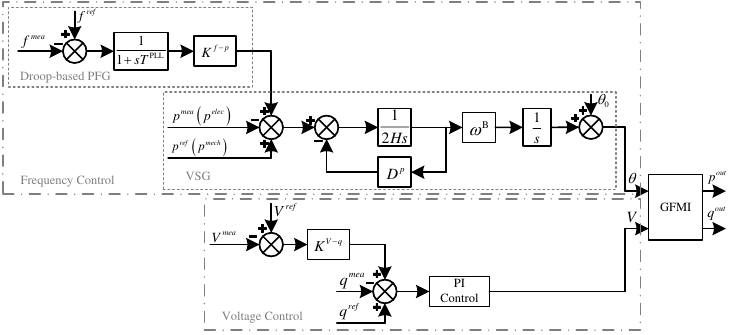}
\caption{Control block diagram of the VSG controlled GFMI-based BESS.}
\label{fig:VSG}
\end{figure}

As shown in Fig.~\ref{fig:VSG}, GFMI-based BESSs regulate MG frequency and voltage using frequency and voltage controllers.

With the VSG, at the core of the frequency control loop, a droop-based PFG is employed to further narrow the frequency deviation after load pick-up events~\cite{Liu2024}. The PFG models the time delay of the phase-locked loop (PLL) while measuring the frequency. Given this structure, for each BESS at $k\in\mathcal{V}$, the transfer function between the variation of the MG frequency $\Delta f_{k}$ and the change of the electrical load power $\Delta p^{elec}_{k}$ is expressed as:
\begin{subequations}\label{eq:frequencyexpLaplace}
\begin{align}
    \Delta f_{k} &= \left[\frac{1 + s T^{\mathrm{PLL}}_k}{s^2 + 2\xi_k\omega^n_{k}s + (\omega^n_k)^2}\right]\frac{\left(\omega^n_k\right)^2\Delta p^{elec}_{k}}{D^p_k + K^{f-p}_k},\\
    \text{where, }&(\omega^n_k)^2 = \frac{D^p_k + K^{f-p}_k}{2H_k T^{\mathrm{PLL}}_k}, \text{ and}\\
    &\xi_k = \frac{2H_k + D^{p}_k T^{\mathrm{PLL}}_k}{2\left(D^{p}_k + K^{f-p}_k\right)}\omega^n_k
\end{align}
\end{subequations}
Here, $H_k$ and $D^p_k$ are the VSG controller's inertia constant and damping factor, $K^{f-p}_k$ is the droop gain of the droop-based PFG, and $T^{\mathrm{PLL}}_k$ is the time constant of the PLL in the MG $k$. $s$ is a Laplace complex frequency. This represents a second-order frequency response whose natural frequency and damping ratio are $\omega^n_k$ and $\xi_k$, respectively. Note that all the parameters in Eq.~\eqref{eq:frequencyexpLaplace} are defined per unit except for $H_k$ and $T^{\mathrm{PLL}}_k$.

The voltage controller adjusts the terminal voltage to follow the defined V-Q droop ($K^{v-q}$). Since the location of the BESS is an optimization variable, all three-phase buses in MG $k$ are considered as potential sites for installation, represented by the subset $\mathcal{B}_{k,3\phi}$. To comply with the format of the
branch power flow of DS, for all $k\in\mathcal{V}$, $n\in\Phi$, and $t\in\mathcal{T}$, the voltage of the GFMI-based BESS at bus $i\in\mathcal{B}_{k,3\phi}$ is controlled as:
\begin{subequations}\label{eq:VSGvoltage}
\begin{align}
    v_{i,n,t,o} \le y^{\mathrm{BESS}}_i\left[\left(V^{b}_i\right)^2 + \Delta v_{i,t,o}^{inc}\right] + (1 - y^{\mathrm{BESS}}_i)M,\\
    v_{i,n,t,o} \ge y^{\mathrm{BESS}}_i\left[\left(V^{b}_i\right)^2 + \Delta v_{i,t,o}^{inc}\right] - (1 - y^{\mathrm{BESS}}_i)M.
\end{align}
\end{subequations}
where $v_{i,n,t,o}$ is the square of the voltage magnitude at bus $i$, phase $n$ and time $t$ in scenario $o$, $V^{b}_i$ is the nominal voltage of bus $i$, $\Delta v_{i,t,o}^{inc}$ is the incremental adjustment of voltage magnitude at bus $i$, $y^{\mathrm{BESS}}_i$ is a binary variable to represent whether the BESS should be installed at bus $i$ or not, and $M$ is a big positive number. Further details about the derivation of the expression of the voltage controller are discussed in Appendix~\ref{ap:VSG}.
%Further details about the derivation of the expressions of the frequency transfer function and the voltage controller are discussed in Appendix~\ref{ap:VSG}.
\subsubsection{Analytic Expressions of Dynamic Frequency Indices}
\begin{figure}[htbp]
\centering
\includegraphics[width=0.48\textwidth]{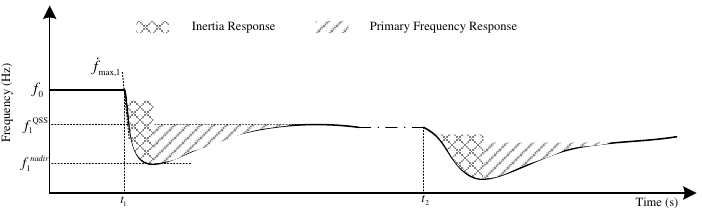}
\caption{Typical frequency response curve.}
\label{fig:FRC}
\end{figure}

As shown in Fig.~\ref{fig:FRC}, when there is a load pick-up event, for example, in time $t_1$, the frequency of the GFMI will dynamically respond according to the VSG and PFG's characteristics. Three significant indices can capture this transient process: the RoCoF $ f^{\mathrm{RoC}}_{\max}$, the quasi-steady state (QSS) frequency $f^{\mathrm{QSS}}$, and the frequency nadir $f^{nadir}$, separately~\cite{Zhang2022}.

During the normal operation period, the frequency of the MG $k\in\mathcal{V}$ is inherited from its QSS frequency at the last time, which implies for all $o\in\mathcal{O}$,
\begin{equation}\label{eq:frequencyBESSori}
f_{k,t,o} = f^{\mathrm{QSS}}_{k,t,o}, \forall t \in \mathcal{T} \setminus (\mathcal{T}^{0} \cup \mathcal{T}^{sync}),
\end{equation}
where $f_{k,t,o}$ is the frequency of the MG $k$, $\mathcal{T}^{0}$ and $\mathcal{T}^{sync}$ are the sets including the BS beginning time and synchronization moments, individually.

For the established VSG-controlled, GFMI-based BESS with the second-order format in the MG $k$, those dynamic frequency indices at the disturbance moment $t\in\mathcal{T}$ for scenario $o\in\mathcal{O}$ can be expressed as follows,
\begin{subequations}\label{eq:frequencyindices}
    \begin{align}
        f_{k,t,o,\max}^{\mathrm{RoC}} &= \frac{-\Delta p^{\mathrm{BESS}}_{k,t,o}}{2H_{k}S^{\mathrm{BESS}}_{k,nom}},\\
        f^{\mathrm{QSS}}_{k,t,o} &= f_{k, t - 1,o} - \frac{\Delta p^{\mathrm{BESS}}_{k,t,o}}{\left(D^p_k + K^{f-p}_k\right)S^{\mathrm{BESS}}_{k,nom}},\\
        f^{nadir}_{k,t,o} &= f_{k,t - 1,o} - \frac{\Delta p^{\mathrm{BESS}}_{k,t,o} \left(1 + \lambda^{nadir}_k\right)}{\left(D^p_k + K^{f-p}_k\right)S^{\mathrm{BESS}}_{k,nom}},\\
        \lambda^{nadir}_k &= \alpha^{nadir}_k\sqrt{1 - \xi^2_k}e^{-\xi_k \omega^n_k t^{nadir}_k},\\
        \alpha^{nadir}_k &= \sqrt{\frac{1 - 2T^{\mathrm{PLL}}_k\xi_k\omega^n_k + \left(T^{\mathrm{PLL}}_k\omega^n_k\right)^2}{1 - \xi^2_k}},\\
        t^{nadir}_k &= \frac{\arctan\left(\frac{\omega^{r}_k T^{\mathrm{PLL}}_k}{\xi_k\omega^{n}_k T^{\mathrm{PLL}}_k - 1}\right)}{\omega^r_k}, \omega^r_k = \omega^n_k\sqrt{1-\xi^2_k},
    \end{align}
\end{subequations}
where $\Delta p^{\mathrm{BESS}}_{k,t,o}$ is the total variation of the output of the BESS in MG $k$ at time $t$ versus time $t-1$, which is equal to the $\Delta p^{elec}_{k,t,o}$. $S^{\mathrm{BESS}}_{k,nom}$ is the nominal rated power of the BESS in the MG $k$, $\lambda^{nadir}_k$ is the frequency nadir ratio versus the QSS frequency, $\alpha^{nadir}_k$ is an intermediate parameter for the frequency nadir, $t^{nadir}_k$ is when the frequency reaches to the nadir after the transient process starts. All these three later parameters related to the frequency nadir are constant parameters. 

\subsection{Modeling of CLPU}\label{sse:clpu}
During BS, initial current demand can surge to several times the nominal level due to simultaneous activation of thermal and motor loads. This phenomenon is known as the CLPU effect~\cite{Li2022}. During the CLPU effect, the load current declines exponentially as the loads gradually diversify over time. The duration of this process can vary from minutes to hours.

In addition, the diversified loads in the DS can be classified as the critical load (CL) and the non-critical load (NL) based on their recovery priority. The CLs in a segment must be restored once the power is delivered to this bus block, while the NLs can be restored in an optimal way.

To catch the dynamic process of the CLPU effect and the different recovery properties of the CL and the NL, two staircase-based functions are developed to model the loads. The buses connected with CLs and NLs are collected into $\mathcal{B}^{\mathrm{CL}}$ and $\mathcal{B}^{\mathrm{NL}}$, respectively. For the segment $m\in\mathcal{G}$, the set storing its buses can be noted as $\mathcal{B}_{m}$. Then, the set including buses in segment $m$ and connected with CLs is defined as  
$\mathcal{B}^{\mathrm{CL}}_m = \mathcal{B}^{\mathrm{CL}} \cap \mathcal{B}_m$. For each bus $i\in\mathcal{B}^{\mathrm{CL}}_m$, $t\in\mathcal{T}$, and $o\in\mathcal{O}$, the CLs during the BS can be modeled as follows,
\begin{subequations}\label{eq:clpucl}
\begin{align}
    \boldsymbol{p}_{i,t,o}^{\mathrm{CL}} = &\;\boldsymbol{p}^{\mathrm{DL}}_{i,t,o}\left[\textstyle\sum_{j=1}^{3}\left(\beta_j \Delta u^{\mathrm{SG}}_{m,t-(j-1),o}\right) + u^{\mathrm{SG}}_{m,t,o}\right],\\
    \boldsymbol{q}_{i,t,o}^{\mathrm{CL}} = &\;\boldsymbol{p}_{i,t,o}^{\mathrm{CL}}\tan\left(\theta^{\mathrm{CL}}_i\right),
\end{align}
\end{subequations}
where $\boldsymbol{p}^{\mathrm{CL}}_{i,t,o}$ and $\boldsymbol{p}^{\mathrm{DL}}_{i,t,o}$ are column vectors storing the restored and nominal active load demand at bus $i$, time $t$, and scenario $o$, separately. $\beta_1$, $\beta_2$, and $\beta_3$ are the CLPU coefficients. $u^{\mathrm{SG}}_{m,t,o}$ is a binary variable representing the energized status of the segment $m$, which highlights the priority of the CL in a segment. $\boldsymbol{q}^{\mathrm{CL}}_{i,t,o}$ and $\boldsymbol{q}^{\mathrm{DL}}_{i,t,o}$ are column vectors storing the restored reactive CL demand and actual reactive load demand at bus $i$, time $t$, and scenario $o$, individually. $\theta^{\mathrm{CL}}_i$ is the power factor angle of the CL at bus $i$.

Similarly, the set including buses in segment $m$ and connected with NLs is defined as $\mathcal{B}^{\mathrm{NL}}_{m} = \mathcal{B}^{\mathrm{NL}} \cap \mathcal{B}_m$. For each bus $i\in\mathcal{B}^{\mathrm{NL}}_{m}$, $t\in\mathcal{T}$, and $o\in\mathcal{O}$, the NLs is modeled as,
\begin{subequations}\label{eq:clpunl}
\begin{align}
    \boldsymbol{p}_{i,t,o}^{\mathrm{NL}} = &\;\boldsymbol{p}^{\mathrm{DL}}_{i,t,o}\left[\textstyle\sum_{j=1}^{3}\left(\beta_j \Delta z^{\mathrm{B}}_{i,t-(j-1),o}\right) + z^{\mathrm{B}}_{i,t,o}\right],\\
    \boldsymbol{q}_{i,t,o}^{\mathrm{NL}} = &\;\boldsymbol{p}_{i,t,o}^{\mathrm{NL}}\tan\left(\theta^{\mathrm{NL}}_i\right),\\
    z^{\mathrm{B}}_{i,t,o} \le &\; u^{\mathrm{SG}}_{m,t,o}, z^{\mathrm{B}}_{i,t - 1,o}\le z^{\mathrm{B}}_{i,t,o},
\end{align}
\end{subequations}
where $\boldsymbol{p}^{\mathrm{NL}}_{i,t,o}$ and $\boldsymbol{q}^{\mathrm{NL}}_{i,t,o}$ are column vectors storing the restored active and reactive NL demand at bus $i$, time $t$, and scenario $o$, separately. $z^{\mathrm{B}}_{i,t,o}$ is a binary variable representing the energized status of the NL at bus $i$, time $t$, and scenario $o$, which is dominated by the segment status. $\theta^{\mathrm{NL}}_i$ is the power factor angle of the NL at bus $i$.
\subsection{Modeling of the SSW}
Before the TG is available, the segments containing the GFMI-based BESS form isolated MGs separately to restore the neighboring segments without self-starting capability. Merging those isolated MGs dynamically during the BS process further improves the system's overall recovery performance. The SSW, capable of telemetering and remote control, can connect the isolated MGs when the synchronization conditions, i.e., identical frequency, voltage magnitude, phase order, and voltage angle, are satisfied. As described in~\cite{Maharjan2024}, these three later conditions combined with the branch flow equations imply no power exchange through the SSW at its action moment. Hence, the synchronizing criteria for closing the SSW can be reformulated as the identical frequency and zero branch flow requirements at the synchronization time.

% a binary variable $\sigma^{\mathrm{SSW}}_{ij,t,o}$ is defined to describe the synchronization action happening between the MG $k$ and $l$ through the SSW $(i,j)$,
Given the set storing the location of switches as $\mathcal{L}^{\mathrm{SW}}$, for each $(i,j)\in\mathcal{L}^{\mathrm{SW}}$, a binary variable $u^{\mathrm{SSW}}_{ij,t,o}$ is defined to describe the status of the SSW $(i,j)$, which is dynamically decided during the optimization process. Hence, the frequency of the GFMI-based BESS in each MG $k\in\mathcal{V}$, which rules other frequencies in other places of the MG it forms, shown in Eq.~\eqref{eq:frequencyBESSori}, can be modified to receive the synchronization adjustment signal from the SSW as follows,
%\begin{subequations}\label{eq:frequencyexpression}
%    \begin{align}
%        f_{k,t,o} &= y^{\mathrm{BESS}}_{k} f_{k,0,o}, \forall t\in\mathcal{T}^0,\\
%        f_{k,t,o} &= f^{\mathrm{QSS}}_{k,t,o} + \sum_{(k,l)\in\mathcal{W}}\Delta u^{syn}_{kl,t,o} \delta^{syn}_{k,t,o}, \forall t\in\mathcal{T} \setminus \mathcal{T}^{0},
%    \end{align}
%\end{subequations}
\begin{subequations}\label{eq:frequencyexpression}
    \begin{align}
        f_{k,t,o} &= y^{\mathrm{BESS}}_{k} f_{k,0,o}, \forall t\in\mathcal{T}^0,\\
        f_{k,t,o} &= f^{\mathrm{QSS}}_{k,t,o} + \delta^{syn}_{k,t,o}\Delta f^{syn}_{k,t,o}, \forall t\in\mathcal{T} \setminus \mathcal{T}^{0},\\
        \delta^{syn}_{k,t,o} &= \textstyle\sum_{(i,j)\in\mathcal{L}^{\mathrm{SW}}}\Delta u^{\mathrm{SSW}}_{ij,t,o}\textstyle\sum_{\{k,l\}\in\mathcal{W}}\Delta u^{syn}_{kl,t,o},
    \end{align}
\end{subequations}
where $y^{\mathrm{BESS}}_{k}$ is a binary variable to present whether the BESS is installed at segment $k$ to form an MG to restore the DS, $f_{k,0,o}$ is the value of the frequency of the BESS in the MG $k$ at the beginning of the BS in scenario $o$, $\delta^{syn}_{k,t,o}$ is the synchronization signal sent from SSW to the MG $k$, $\Delta f^{syn}_{k,t,o}$ is the synchronization frequency adjustment amount of the MG $k$, and $u^{syn}_{kl,t,o}$ is a binary variable representing the synchronization status between the MG $k$ and $l$.

In addition, there is no power flow on the SSW $(i,j)$ when it closes to synchronize two MGs, given the fact that the voltage magnitudes and angles are the same at each end of the SSW at that moment, which reveals for all $(i,j)\in\mathcal{L}^{\mathrm{SW}}$ and $t\in\mathcal{T}$,
%\begin{subequations}\label{eq:sswnobranchflow}
%    \begin{align}
%    \left(\sigma^{\mathrm{SSW}}_{ij,t,o} - 1\right)\boldsymbol{M}_{|\Phi_{ij}|} &\le \boldsymbol{p}_{ij,t,o} \le \left(1 - \sigma^{\mathrm{SSW}}_{ij,t,o}\right)\boldsymbol{M}_{|\Phi_{ij}|}, \\
%    \left(\sigma^{\mathrm{SSW}}_{ij,t,o} - 1\right)\boldsymbol{M}_{|\Phi_{ij}|} &\le \boldsymbol{q}_{ij,t,o} \le \left(1 - \sigma^{\mathrm{SSW}}_{ij,t,o}\right)\boldsymbol{M}_{|\Phi_{ij}|}, 
%    \end{align}
%\end{subequations}
\begin{subequations}\label{eq:sswnobranchflow}
    \begin{align}
    \left(\Delta u^{\mathrm{SSW}}_{ij,t,o} \hspace{-0.5em}-\hspace{-0.2em} 1\right)\boldsymbol{M}_{|\Phi_{ij}|} &\le \boldsymbol{p}_{ij,t,o} \le \left(1 \hspace{-0.2em}-\hspace{-0.3em} \Delta u^{\mathrm{SSW}}_{ij,t,o}\right)\boldsymbol{M}_{|\Phi_{ij}|}, \\
    \left(\Delta u^{\mathrm{SSW}}_{ij,t,o} \hspace{-0.5em}-\hspace{-0.2em} 1\right)\boldsymbol{M}_{|\Phi_{ij}|} &\le \boldsymbol{q}_{ij,t,o} \le \left(1 \hspace{-0.2em}-\hspace{-0.3em} \Delta u^{\mathrm{SSW}}_{ij,t,o}\right)\boldsymbol{M}_{|\Phi_{ij}|}, 
    \end{align}
\end{subequations}
where $\Phi_{ij}$ is a set including the phases of branch $ij$, $|\cdot|$ is an operator calculating the cardinality of a set, and $\boldsymbol{M}$ is a column vector with $|\Phi_{ij}|$ big positive number. $\boldsymbol{p}_{ij,t,o}$ and $\boldsymbol{q}_{ij,t,o}$ are the column vectors storing the branch active and reactive power at switch $ij$, respectively.

Further explanations of the SSW about the BS action are described in Section~\ref{sse:secondstagebs}.
\vspace{0em}
\section{Mathematical Formulation}\label{se:3}
In this section, we first describe the resource allocation process for deploying BS equipment at the initial stage. Next, we formulate the SDMG-BS framework for the DS at the second stage. Finally, we present a comprehensive narration of this two-stage stochastic model.
% \subsection{Uncertainty Modeling}
\vspace{-1em}
\subsection{First Stage Resource Allocation Problem}
\subsubsection{Objective of Resource Allocation}
The objective of the resource allocation stage is to minimize the total investment cost of BS resources, by optimizing the number and size of GFMI-based BESSs:
\begin{equation}
    \min F^{alloc} = \textstyle\sum_{k\in\mathcal{V}}\left(y^{\mathrm{BESS}}_{k} + S^{\mathrm{BESS}}_{k,nom} + E^{\mathrm{BESS}}_{k,nom}\right),
\end{equation}
where $F^{alloc}$ is the allocation objective, and $E^{\mathrm{BESS}}_{k,nom}$ denotes the nominal rated capacity of the BESS for MG $k\in\mathcal{V}$.
\subsubsection{Constraints on BESS Allocation inside MGs}
For each MG $k\in\mathcal{V}$, the installed GFMI-based BESS at its three-phase buses, satisfying,
\begin{subequations}\label{eq:allocationinsideMG}
\begin{align}
        y_{i}^{\mathrm{BESS}}S_{\min}^{\mathrm{BESS}} &\le S_{i,nom}^{\mathrm{BESS}} \le y_{i}^{\mathrm{BESS}}S_{\max}^{\mathrm{BESS}},\\
        y_{i}^{\mathrm{BESS}}E_{\min}^{\mathrm{BESS}} &\le E_{i,nom}^{\mathrm{BESS}} \le y_{i}^{\mathrm{BESS}}E_{\max}^{\mathrm{BESS}},
\end{align}
\end{subequations}
%\begin{equation}\label{eq:allocationinsideMG}
%    y_{i}^{\mathrm{BESS}}\chi_{\min}^{\mathrm{BESS}} \le \chi_{i,nom}^{\mathrm{BESS}} \le y_{i}^{\mathrm{BESS}}\chi_{\max}^{\mathrm{BESS}}, \chi\in\{S, E\},
%\end{equation}
where $S_{\min}^{\mathrm{BESS}}$, $S_{\max}^{\mathrm{BESS}}$ are the minimum and maximum allowable rated powers, and $E_{\min}^{\mathrm{BESS}}$, $E_{\max}^{\mathrm{BESS}}$ are the minimum and maximum allowable capacities for a single BESS. The nominal capacity $E_{i,nom}^{\mathrm{BESS}}$ pertains to the BESS at bus $i\in\mathcal{B}_{k,3\phi}$.

Additionally, each MG requires exactly one GFMI-based BESS to provide frequency and voltage references before merging with other MGs:
\begin{subequations}\label{eq:bessrainsideMGp2}
    \begin{align}
        \textstyle\sum_{i\in\mathcal{B}_{k,3\phi}}y_{i}^{\mathrm{BESS}} &= y_{k}^{\mathrm{BESS}},\\
        \textstyle\sum_{i\in\mathcal{B}_{k,3\phi}}\chi^{\mathrm{BESS}}_{i,nom} &= \chi^{\mathrm{BESS}}_{k,nom}, \chi\in\{S, E\},\label{eq:bessrainsideMGp2b}
    \end{align}
\end{subequations}
while the Eq.~\eqref{eq:bessrainsideMGp2b} is stated to simplify the expression of the formulation.
\subsubsection{Constraints on BESS Allocation among MGs}
To ensure the BS process is successful, at least one MG should be available before the TG is available. In addition, the total deployed BESS resources for the BS are restricted by the DS budget. These requirements among the MGs can be concluded as,
%\begin{subequations}
%    \begin{align}
%        \sum_{k\in\mathcal{V}}y_{k}^{\mathrm{BESS}} &\ge 1,\\
%        \sum_{k\in\mathcal{V}}S^{\mathrm{BESS}}_{k,nom} &\le S^{\mathrm{BESS}}_{budget},\\
%        \sum_{k\in\mathcal{V}}E^{\mathrm{BESS}}_{k,nom} &\le E^{\mathrm{BESS}}_{budget},
%    \end{align}
%\end{subequations}
\begin{subequations}
    \begin{align}
        \textstyle\sum_{k\in\mathcal{V}}y_{k}^{\mathrm{BESS}} &\ge 1,\\
        \textstyle\sum_{k\in\mathcal{V}}\chi^{\mathrm{BESS}}_{k,nom} &\le \chi^{\mathrm{BESS}}_{budget}, \chi\in\{S, E\},
    \end{align}
\end{subequations}
where $S^{\mathrm{BESS}}_{budget}$ and $E^{\mathrm{BESS}}_{budget}$ are the overall budgets for the rated power and capacity of the BESS assigned to the BS, respectively.
\subsubsection{Constraints on SSW Allocation}
Some ESWs are replaced with SSWs to enable synchronization during BS. To minimize costs and maximize flexibility, the number of deployed SSWs is constrained as,
\begin{equation}\label{eq:sswallocation}
    \textstyle\sum_{(i,j)\in\mathcal{L}^{\mathrm{SW}}}y_{ij}^{\mathrm{SSW}} = \textstyle\sum_{k\in\mathcal{V}} y^{\mathrm{BESS}}_k - 1,
\end{equation}
where $y_{ij}^{\mathrm{SSW}}$ is a binary variable to present whether the switch at $(i,j)$ is replaced with a SSW.
\subsection{Second Stage SDMG-BS Problem}\label{sse:secondstagebs}
\subsubsection{Objective of BS}
The second-stage objective is to maximize the total restored loads over the restoration horizon,
\begin{align}
    \max F^{restor}_o = &\textstyle\sum_{t\in\mathcal{T}}\left[\gamma^{\mathrm{CL}}\textstyle\sum_{i\in\mathcal{B}^{\mathrm{CL}}} \left(\boldsymbol{1}^{T}_{|\Phi_i|}\boldsymbol{p}^{\mathrm{CL}}_{i,t,o}\right)\notag\right.\\
    &\left.+ \gamma^{\mathrm{NL}}\textstyle\sum_{i\in\mathcal{B}^{\mathrm{NL}}}\left(\boldsymbol{1}^{T}_{|\Phi_i|}\boldsymbol{p}^{\mathrm{NL}}_{i,t,o}\right)\right]\Delta t,
\end{align}
where $F^{restor}_o$ is the objective function of the BS in scenario $o$, $\gamma^{\mathrm{CL}}$ and $\gamma^{\mathrm{NL}}$ are weighting coefficients for the CL and NL, separately. $\Phi_i$ is the set storing the phases of loads connected at bus $i$, $\boldsymbol{1}_{|\Phi_i|}^{T}$ is the transpose of a column vector with all $|\Phi_i|$ elements being one, $\Delta t$ is the duration between $t$ and $t-1$.
\subsubsection{Constraints on Switches Actions}
\paragraph{Energized switches (ESW) action}
For each switch $(i,j)\in\mathcal{L}^{\mathrm{SW}}$, if it's not chosen to be replaced with the SSW, it operates as an ESW during the BS, whose action is constrained as, 
\begin{subequations}\label{eq:eswaction}
    \begin{align}
    u^{\mathrm{ESW}}_{ij,t,o} &\le 1 - y^{\mathrm{SSW}}_{ij},\label{eq:eswaction1}\\
    u^{\mathrm{ESW}}_{ij,t,o} &\le u^{\mathrm{B}}_{i,t - 1,o} + u^{\mathrm{B}}_{j,t - 1,o},\label{eq:eswaction2}\\
    \Delta u^{\mathrm{ESW}}_{ij,t,o} &\le 2 - u^{\mathrm{B}}_{i,t - 1,o} - u^{\mathrm{B}}_{j,t - 1,o},\label{eq:eswaction3}
    \end{align}
\end{subequations}
where $u^{\mathrm{ESW}}_{ij,t,o}$ is a binary variable to indicate whether the ESW $(i,j)$ is closed, $u^{\mathrm{B}}_{i,t,o}$ is a binary variable to present the energized status of the bus $i$ of the ESW $(i,j)$. As shown in Eq.~\eqref{eq:eswaction1}, the action of the ESW $(i,j)$ will be blocked if the SSW is placed here. The ESW $(i,j)$ can act to energize other non-MG segments once one end of it belongs to a powered MG, but it cannot be used to synchronize two operated MGs as depicted in Eq.~\eqref{eq:eswaction2} and~\eqref{eq:eswaction3}.

Moreover, once a ESW $(i,j)\in\mathcal{L}^{\mathrm{SW}}$ is closed to pick up a non-MG segment, the frequency of the MG it is associated with is passed to that new energized segment as follows,
%\begin{subequations}\label{eq:eswfrequency}
%    \begin{align}
%        f_{i,t,o} &\le f_{j,t,o} + \left(1 - u^{\mathrm{ESW}}_{ij,t,o}\right)M,\\
%        f_{i,t,o} &\ge f_{j,t,o} - \left(1 - u^{\mathrm{ESW}}_{ij,t,o}\right)M,
%    \end{align}
%\end{subequations}
\begin{equation}\label{eq:eswfrequency}
        f_{j,t,o} \hspace{-0.2em}-\hspace{-0.2em} \left(1 \hspace{-0.2em}-\hspace{-0.2em}u^{\mathrm{ESW}}_{ij,t,o}\right)M \le f_{i,t,o} \le f_{j,t,o} + \left(1\hspace{-0.2em} - \hspace{-0.2em}u^{\mathrm{ESW}}_{ij,t,o}\right)M,
\end{equation}
where $f_{i,t,o}$ is the bus $i$'s frequency.
\paragraph{SSWs action}
If a switch $(i,j)\in\mathcal{L}^{\mathrm{SW}}$ is chosen to be replaced with a SSW, it will be capable of synchronizing two MGs during the BS, whose action is constrained as,
\begin{subequations}\label{eq:sswaction}
    \begin{align}
    u^{\mathrm{SSW}}_{ij,t,o} &\le y^{\mathrm{SSW}}_{ij},\label{eq:sswaction1}\\
    u^{\mathrm{SSW}}_{ij,t,o} &\ge u^{\mathrm{SSW}}_{ij,t - 1,o},\label{eq:sswaction2}\\
    2u^{\mathrm{SSW}}_{ij,t,o} &\le u^{\mathrm{B}}_{i,t - 1,o} + u^{\mathrm{B}}_{j,t - 1,o},\label{eq:sswaction3}\\
    f_{i,t,o} &\le f_{j,t,o} + \left(1 - u^{\mathrm{SSW}}_{ij,t,o}\right)M + \frac{\epsilon}{2},\label{eq:sswaction4}\\
    f_{i,t,o} &\ge f_{j,t,o} - \left(1 - u^{\mathrm{SSW}}_{ij,t,o}\right)M - \frac{\epsilon}{2},\label{eq:sswaction5}
    \end{align}
\end{subequations}
where $u^{\mathrm{SSW}}_{ij,t,o}$ is a binary variable to indicate whether the SSW $(i,j)$ is closed, and $\epsilon$ is a small positive number. As illustrated in Eq.~\eqref{eq:sswaction1}, the SSW $(i,j)$ will be blocked if it is not deployed. Eq.~\eqref{eq:sswaction2} shows that the action of SSW is irreversible. Eq.~\eqref{eq:sswaction3} requires the SSW $(i,j)$ can only be used to do the synchronization. Different from Eq.~\eqref{eq:eswfrequency}, the SSW can only be closed once the frequencies at both ends of it are identical, as shown in Eq.~\eqref{eq:sswaction4} and~\eqref{eq:sswaction5}.

% In addition, the moment of the synchronization executed by the SSW $(i,j)$ can be captured as follows,
%\begin{equation}\label{eq:sswsynchronizationmoment}
%    \sigma^{\mathrm{SSW}}_{ij,t,o} = \Delta u^{\mathrm{SSW}}_{ij,t,o}\left(u^{\mathrm{B}}_{i,t - 1,o} + u^{\mathrm{B}}_{j,t - 1,o} - u^{\mathrm{SSW}}_{ij,t,o}\right).
%\end{equation}
\subsubsection{Constraints on BS Energization}
\paragraph{Segment energization status}
Given the set storing the non-switchable line of DS as $\mathcal{L}^{\mathrm{LN}}$, and the set collecting the branch inside a segment $m\in\mathcal{G}$ as $\mathcal{L}_m$, the set including the lines inside the segment $m$ can be defined as $\mathcal{L}^{\mathrm{LN}}_m = \mathcal{L}^{\mathrm{LN}} \cap \mathcal{L}_m$. Similarly, the set holding the switch inside a segment can be defined as $\mathcal{L}^{\mathrm{SW}}_m = \mathcal{L}^{\mathrm{SW}} \cap \mathcal{L}_m$. And then, for each segment $m$, its energization status is constrained as follows,
\begin{subequations}\label{eq:segmentstatus}
    \begin{align}
    u^{\mathrm{SG}}_{m,t,o} &\ge u^{\mathrm{SG}}_{m,t - 1,o},\label{eq:segmentstatus1}\\
    u^{\mathrm{SG}}_{m,t,o} &= u^{\mathrm{B}}_{i,t,o}, \forall i\in\mathcal{B}_m,\label{eq:segmentstatus2}\\
    u^{\mathrm{SG}}_{m,t,o} &= u^{\mathrm{L}}_{ij,t,o}, \forall (i,j)\in\mathcal{L}^{\mathrm{LN}}_m,\label{eq:segmentstatus3}\\
    u^{\mathrm{SG}}_{m,t,o} &\ge u^{\mathrm{ESW}}_{ij,t,o}, \forall (i,j)\in\mathcal{L}^{\mathrm{SW}}_m,\label{eq:segmentstatus4}\\
    \textstyle\sum_{(i,j)\in\mathcal{L}^{\mathrm{SW}}_m}\Delta u^{\mathrm{ESW}}_{ij,t,o} &\le u^{\mathrm{SG}}_{m,t - 1,o}M + 1,\label{eq:segmentstatus5}
    \end{align}
\end{subequations}
where $u^{\mathrm{L}}_{ij,t,o}$ is a binary variable that stands for the energized status of the branch $(i,j)$. As described in Eq.~\eqref{eq:segmentstatus1}, once the segment is energized, it will be powered until the end of the optimization. Meanwhile, all the buses and non-switchable lines inside the segment should have the same status as the segment itself, as shown in Eq.~\eqref{eq:segmentstatus2} and~\eqref{eq:segmentstatus3}. The black-out segment can only be energized by one of ESWs belonging to it at the same time, as shown in Eq.~\eqref{eq:segmentstatus4} and~\eqref{eq:segmentstatus5}.
\paragraph{Radiality of the DS}
The radial topology of DS is required to maintain during the BS, which is dynamically  constrained as follows,
\begin{subequations}\label{eq:radiality}
    \begin{align}
    &\sum_{\left(i,j\right)\in\mathcal{L}}u^{\mathrm{L}}_{ij,t,o} = \sum_{i\in\mathcal{B}}u^{\mathrm{B}}_{i,t,o} - R_{t,o},\label{eq:radiality1}\\
    & R_{t,o} = \sum_{k\in\mathcal{V}}y^{\mathrm{BESS}}_{k} + \sum_{i\in\mathcal{B}^{\mathrm{TG}}}u^{\mathrm{B}}_{i,t,o} - \sum_{(i,j)\in\mathcal{L}^\mathrm{SW}}u^{\mathrm{SSW}}_{ij,t,o},\label{eq:radiality2}
    \end{align}
\end{subequations}
where $R_{t,o}$ is the number of root buses, $\mathcal{B}^{\mathrm{TG}}$ is the set collecting the TG buses. The radiality of DS is guaranteed by Eq.~\eqref{eq:radiality1}, while $R_{t,o}$ is dynamically decided during the BS process in Eq.~\eqref{eq:radiality2}.
\paragraph{TG outage and recovery}
The TG is unavailable during the BS until the fault transmission line is repaired. Given the set storing the TG bus as $\mathcal{B}^{\mathrm{TG}}$, the output and status of TG $i\in\mathcal{B}^{\mathrm{TG}}$ is constrained with its indicative binary variable $u^{\mathrm{TG}}_{i, t, o}$ at each $t\in\mathcal{T}$ and $o\in\mathcal{O}$ as follows,
\begin{subequations}
    \begin{align}
    &\max_{n\in\Phi}\left\{\left(p^{\mathrm{TG}}_{i,n,t,o}\right)^2 +\left(q^{\mathrm{TG}}_{i,n,t,o})^2\right)\right\} \le \left(\tfrac{1}{3}S^{\mathrm{TG}}_{i, max}\right)^2,\\
    &u^{\mathrm{TG}}_{i,t,o}\boldsymbol{1}_{|\Phi|} \le \boldsymbol{v}_{i,t,o} \le u^{\mathrm{TG}}_{i,t,o}\boldsymbol{1}_{|\Phi|},\\ 
    &60u^{\mathrm{TG}}_{i,t,o} \le f^{\mathrm{TG}}_{i, t, o} \le 60u^{\mathrm{TG}}_{i,t,o},
    \end{align}
where $p^{\mathrm{TG}}_{i,n,t,o}$ and $q^{\mathrm{TG}}_{i,n,t,o}$ are the active and reactive output of TG's nth phase. $S^{\mathrm{TG}}_{i, max}$ is the maximum rated power of TG at bus $i$. $f^{\mathrm{TG}}_{i, t, o}$ is the TG's frequency.
\end{subequations}
\subsubsection{Constraints on Synchronization Processes}
Define the set, including buses at of switches belonging to segment $k$, as $\mathcal{B}^{\mathrm{SW}}_k$. Then, for each $k\in\mathcal{V}$, the frequencies of bus $i\in\mathcal{B}^{\mathrm{SW}}_k$ inside the MG $k$ are ruled as follows,
\begin{subequations}\label{eq:propagatefrequency1}
    \begin{align}
        f_{i,t,o} \le f_{k,t,o} + (1 - y^{\mathrm{BESS}}_k)M,\\
        f_{i,t,o} \ge f_{k,t,o} - (1 - y^{\mathrm{BESS}}_k)M.
    \end{align}
\end{subequations}

Furthermore, for each general segment $m\in\mathcal{G}$, the frequencies of bus $i\in\mathcal{B}^{\mathrm{SW}}_m$ should be uniform at every moment due to their same energized status shown in Eq.~\eqref{eq:segmentstatus2}, which is concluded as follows,
\begin{equation}\label{eq:propagatefrequency2}
    f_{i,t,o} = f_{j,t,o}, \forall j\in\mathcal{B}^{\mathrm{SW}}_m, j\neq i.
\end{equation}

Based on the propagation mechanism of frequency during the energization stage of BS depicted in Eq.~\eqref{eq:propagatefrequency1} and~\eqref{eq:propagatefrequency2}, the synchronization moment happening between the MG $k$ and $l$ in the set $\mathcal{V}$ at time $t\in\mathcal{T}$ in scenario $o\in\mathcal{O}$, where $\{k,l\}\in\mathcal{W}$, can be captured as,
\begin{subequations}
    \begin{align}
        u^{syn}_{kl,t,o} &\ge u^{syn}_{kl,t - 1,o},\\
        (2 - y^{\mathrm{BESS}}_k - y^{\mathrm{BESS}}_l) &= \mu_{kl},\\
        u^{syn_{-}}_{kl,t,o} + u^{syn}_{kl,t,o} + u^{syn_{+}}_{kl,t,o} &= 1,\\
        \epsilon u^{syn_{-}}_{kl,t,o} - \frac{\epsilon}{2}u^{syn}_{kl,t,o} - M u^{syn_{+}}_{kl,t,o} &\le f_{l,t,o} - f_{k,t,o} + \mu_{kl},\\
        M u^{syn_{-}}_{kl,t,o} + \frac{\epsilon}{2}u^{syn}_{kl,t,o} - \epsilon u^{syn_{+}}_{kl,t,o} &\ge f_{l,t,o} - f_{k,t,o} + \mu_{kl},
    \end{align}
\end{subequations}
where $\mu_{kl}$ is an intermediate variable standing for whether deploying GFMI-based BESS inside the segment $k$ and $l$ at the same time. $u^{syn_{-}}_{kl,t,o}$ and $u^{syn_{+}}_{kl,t,o}$ are the binary variables representing the left and right synchronization indicators between the MG $k$ and $l$, respectively.

Additionally, to decrease the inrush current of the GFMI-based BESS at the synchronization moment, the number of MGs merged into another MG should be restricted to one. For any two elements $\{k,l\}$ and $\{m,n\}$ in $\mathcal{W}$, define the set $\mathcal{X} = \{\{k,l,m,n\}\}$, where $|\{k,l\}\cap\{m,n\}| = 1$. And then, for each $\{k,l,m,n\}\in\mathcal{X}$, the above extra synchronization constraint is modeled as,
\begin{equation}\label{eq:synchronizationnumberofmergedMGs}
    \Delta u^{syn}_{kl,t,o} + \Delta u^{syn}_{mn,t,o} - u^{syn}_{ln,t,o} \le 1,
\end{equation}
where $\{l,n\} = \left(\{k,l\}\cup\{m,n\}\right)\setminus\left(\{k,l\}\cap\{m,n\}\right)$.
\subsubsection{Constraints on Frequency Security}
Given the dynamic frequency indices of the transient process and the full frequency expression under normal operation in Eq.~\eqref{eq:frequencyindices} and~\eqref{eq:frequencyexpression}, respectively, for each $k\in\mathcal{V}$ and $t\in\mathcal{T}$, the following constraints are introduced to guarantee the frequency security during the transient and normal process of the BS,
\begin{subequations}\label{eq:frequencysecurity}
    \begin{align}
    \lfloor f^{\mathrm{RoC}}_{\max} \rfloor &\le f^{\mathrm{RoC}}_{k,t,o,\max} \le \lceil f^{\mathrm{RoC}}_{\max} \rceil\label{eq:frequencysecurityRoC},\\
    \lfloor f^{\mathrm{QSS}} \rfloor &\le f^{\mathrm{QSS}}_{k,t,o} \le \lceil f^{\mathrm{QSS}} \rceil\label{eq:frequencysecurityQSS},\\
    \lfloor f^{\mathrm{nadir}} \rfloor &\le f^{\mathrm{nadir}}_{k,t,o} \le \lceil f^{\mathrm{nadir}}\rceil\label{eq:frequencysecuritynadir},\\
    \lfloor f_k \rfloor &\le f_{k,t,o} \le \lceil f_k \rceil\label{eq:frequencysecurityfrequency},
    \end{align}
\end{subequations}
where $\lfloor \cdot \rfloor$ and $\lceil \cdot \rceil$ are the lower and upper limit operators of a variable, respectively.
\subsubsection{Constraints on Unbalanced Linear Power Flow}
Define the set collecting the child buses of bus $i$ as $\mathcal{B}_i^{ch}$, the nodal power balance constraints for all $i\in\mathcal{B}$ at time $t\in\mathcal{T}$ in scenario $o\in\mathcal{O}$ can be expressed as follow,
\begin{subequations}\label{eq:power_flow_unbalanced}
   \begin{align}   
    \boldsymbol{p}_{i,t,o} &= \sum_{j\in \mathcal{B}_i^{ch}}\left(\boldsymbol{\Lambda}_{|\Phi_{hi}|\times|\Phi_{ij}|}\boldsymbol{p}_{ij,t,o}\right) - \boldsymbol{p}_{hi,t,o}\\
    \boldsymbol{q}_{i,t,o} &= \sum_{j\in \mathcal{B}_i^{ch}}\left(\boldsymbol{\Lambda}_{|\Phi_{hi}|\times|\Phi_{ij}|}\boldsymbol{q}_{ij,t,o}\right) - \boldsymbol{q}_{hi,t,o},\\
    \boldsymbol{p}_{i,t,o} &= \boldsymbol{p}^{\mathrm{TG}}_{i,t,o} + \boldsymbol{p}^{\mathrm{BESS}}_{i,t,o} + \boldsymbol{p}^{\mathrm{PV}}_{i,t,o} - \boldsymbol{p}^{\mathrm{CL}}_{i,t,o} - \boldsymbol{p}^{\mathrm{NL}}_{i,t,o}\\
    \boldsymbol{q}_{i,t,o} &= \boldsymbol{q}^{\mathrm{TG}}_{i,t,o} + \boldsymbol{q}^{\mathrm{BESS}}_{i,t,o} + \boldsymbol{q}^{\mathrm{PV}}_{i,t,o} - \boldsymbol{q}^{\mathrm{CL}}_{i,t,o} - \boldsymbol{q}^{\mathrm{NL}}_{i,t,o}
\end{align}
\end{subequations}
where $\boldsymbol{p}_{i,t,o}$ and $\boldsymbol{q}_{i,t,o}$ are the column vectors storing the injection active and reactive power at bus $i$, individually. $\boldsymbol{\Lambda}_{|\Phi_{hi}|\times|\Phi_{ih}|}$ is a matrix with $|\Phi_{hi}|$ rows and $|\Phi_{ij}|$ columns, whose elements in $\Phi_j$ row all are one. $\boldsymbol{p}^{\mathrm{BESS}}_{i,t,o}$ and $\boldsymbol{q}^{\mathrm{BESS}}_{i,t,o}$ are the column vectors collecting the active and reactive output of the BESS at bus $i$, separately. $\boldsymbol{p}^{\mathrm{PV}}_{i,t,o}$ and $\boldsymbol{q}^{\mathrm{PV}}_{i,t,o}$ are the column vectors collecting the active and reactive output of the PV at bus $i$, respectively.

The phase voltages located at the ends of a branch $(i,j)\in\mathcal{L}$ at time $t\in\mathcal{T}$ in scenario $o\in\mathcal{O}$ are combined with the following constraints,
\begin{subequations}\label{eq:line_flow_unbalanced}
\begin{align}
    \boldsymbol{v}_{j,t,o} \le &\left[\boldsymbol{v}_{i,t,o}^{\Phi_{ij}} - 
    2\left(\boldsymbol{r}_{ij}\boldsymbol{p}_{ij,t,o} + \boldsymbol{x}_{ij}\boldsymbol{q}_{ij,t,o}\right)\right.\notag\\
    &\left.+ \left(1-u^{\mathrm{L}}_{ij,t,o}\right)\boldsymbol{M}^{\Phi_{ij}}\right],\label{eq:line_flow_r}\\
    \boldsymbol{v}_{j,t,o} \ge &\left[\boldsymbol{v}_{i,t,o}^{\Phi_{ij}} - 
    2\left(\boldsymbol{r}_{ij}\boldsymbol{p}_{ij,t,o} + \boldsymbol{x}_{ij}\boldsymbol{q}_{ij,t,o}\right)\right.\notag\\
    &\left.- \left(1-u^{\mathrm{L}}_{ij,t,o}\right)\boldsymbol{M}^{\Phi_{ij}}\right],\label{eq:line_flow_l}
\end{align}
\end{subequations}
with,
\begin{equation}
    u^{\mathrm{L}}_{ij,t} = u^{\mathrm{ESW}}_{ij,t} + u^{\mathrm{SSW}}_{ij,t}, \forall (i,j)\in\mathcal{L}^{\mathrm{SW}},
\end{equation}
where $\boldsymbol{v}_{j,t,o}$ is a column vector storing the square of the voltage magnitude at bus $j$, $\cdot^{\Phi_{ij}}$ is an operator extracting the phase elements of branch $(i,j)$. $\boldsymbol{r}_{ij}$ and $\boldsymbol{x}_{ij}$ are matrices associated with branch resistance and reactance, whose calculations are described in~\cite{Cheng2022}.

Moreover, to keep the normal operation of the restored segments, for each $(i,j)\in\mathcal{L}$ and $i\in\mathcal{B}$, the following constraints need to be satisfied,
\begin{subequations}\label{eq:powerflowsecurity}
    \begin{align}
    -u^{\mathrm{L}}_{ij,t,o}\boldsymbol{M}^{\Phi_{ij}} &\le \boldsymbol{p}_{ij,t,o} \le u^{\mathrm{L}}_{ij,t,o}\boldsymbol{M}^{\Phi_{ij}},\\
    -u^{\mathrm{L}}_{ij,t,o}\boldsymbol{M}^{\Phi_{ij}} &\le \boldsymbol{q}_{ij,t,o} \le u^{\mathrm{L}}_{ij,t,o}\boldsymbol{M}^{\Phi_{ij}},\\  
    u_{i,t,o}^{\mathrm{B}}\lfloor \boldsymbol{v}_{i} \rfloor &\le \boldsymbol{v}_{i,t,o} \le u_{i,t,o}^{\mathrm{B}}\lceil \boldsymbol{v}_{i} \rceil\label{eq:powerflowsecurity3},
    \end{align}
\end{subequations}
\subsubsection{Constraints on Resource Output}
\paragraph{BESS output}
The output of the BESS at bus $i\in\mathcal{B}_{k,3\phi}$ for all $k\in\mathcal{V}$, time $t\in\mathcal{T}$, and scenario $o\in\mathcal{O}$, is constrained by the planned rated power and capacity at the first stage, which is illustrated as follows,
\begin{subequations}\label{eq:BESScontrol}
\begin{align}
    &\max_{n\in\Phi}\left\{\left(p^{\mathrm{BESS}}_{i,n,t,o}\right)^2 +\left(q^{\mathrm{BESS}}_{i,n,t,o})^2\right)\right\} \le \left(\tfrac{1}{3}S^{\mathrm{BESS}}_{i, nom}\right)^2,\\
    &SoC_{i,t,o} = SoC_{i,t-1,o} - \boldsymbol{1}_{|\Phi_i|}^{T}\boldsymbol{p}^{\mathrm{BESS}}_{i,t,o}/E^{\mathrm{BESS}}_{i,nom},\\
    &\lfloor SoC \rfloor \le SoC_{i,t,o} \le \lceil SoC \rceil,\label{eq:SoCconstraint}
\end{align}
\end{subequations}
where $p^{\mathrm{BESS}}_{i,n,t,o}$ and $q^{\mathrm{BESS}}_{i,n,t,o}$ are the $n$-th phase value of $\boldsymbol{p}^{\mathrm{BESS}}_{i,t,o}$ and $\boldsymbol{q}^{\mathrm{BESS}}_{i,t,o}$, separately. $SoC_{i,t,o}$ is the state of charge (SoC) of the BESS at bus $i$.

In addition, the next constraints are introduced to express the active output of the BESS belonging to the MG $k\in\mathcal{V}$,
\begin{equation}
    \boldsymbol{p}^{\mathrm{BESS}}_{k,t,o} = \textstyle\sum_{i\in\mathcal{B}_{k,3\phi}}\boldsymbol{p}^{\mathrm{BESS}}_{i,t,o}.
\end{equation}
\paragraph{PV output}
The behind-meter PV system's active output depends on the fluctuating outer environment, while its reactive output is in a fixed ratio to the former one. Given the PV systems output rate as $\eta_{t,o}$, and the set containing the buses with PVs as $\mathcal{B}^\mathrm{PV}$, the PV system at bus $i\in\mathcal{B}^\mathrm{PV}$ follows the constraints,
\begin{subequations}\label{eq:PVcontrol}
\begin{align}
    \boldsymbol{p}^{\mathrm{PV}}_{i,t,o} &= \tfrac{1}{3}S^{\mathrm{PV}}_{i,nom}\eta_{t,o}\boldsymbol{1}_{|\Phi_i|},\\
    \boldsymbol{q}^{\mathrm{PV}}_{i,t,o} &= 0.352\boldsymbol{p}^{\mathrm{PV}}_{i,t,o},
\end{align}
\end{subequations}
where $S^{\mathrm{PV}}_{i,nom}$ is the nominal rated power of the PV at bus $i$.
\subsection{Full Resource Allocation Formulation in SDMG-BS}
The complete two-stage stochastic resource allocation model integrates the objectives and constraints from the first and second stages as follows,
\begin{align}
    &\min \;\; F^{alloc} - \sum_{o\in\mathcal{O}}\pi_{o}F^{restor}_o,\\
    &\text{s.t.}\;\; \text{Resoure Allocation Constraints: Eq.}~\eqref{eq:allocationinsideMG}\sim\eqref{eq:sswallocation},\\
    &\quad\;\;\,\text{Switches Action Constraints: Eq.}~\eqref{eq:eswaction}\sim\eqref{eq:sswaction},\\
    &\quad\;\;\,\text{Energization Constraints: Eq.}~\eqref{eq:segmentstatus}\sim\eqref{eq:radiality},\\
    &\quad\;\;\,\text{Synchronization Constraints: Eq.}~\eqref{eq:sswnobranchflow},~\eqref{eq:propagatefrequency1}\sim\eqref{eq:synchronizationnumberofmergedMGs},\\
    &\quad\;\;\,\text{Frequency Security Constraints: Eq.}~\eqref{eq:frequencyindices},~\eqref{eq:frequencyexpression},\eqref{eq:frequencysecurity},\\
    &\quad\;\;\,\text{Power Flow Constraints: Eq.}~\eqref{eq:VSGvoltage}~\eqref{eq:power_flow_unbalanced}\sim\eqref{eq:powerflowsecurity},\\
    &\quad\;\;\,\text{CLPU Constraints: Eq.}~\eqref{eq:clpucl}\sim\eqref{eq:clpunl},\\
    &\quad\;\;\,\text{Resource Output Constraints: Eq.}~\eqref{eq:BESScontrol}\sim\eqref{eq:PVcontrol},
\end{align}
where $\pi_{o}$ is the probability of the scenario $o$.
%\section{Solution Methodology}\label{se:4}
\section{Simulation Setup}\label{se:4}
In this section, we first outline the adopted simulation system and parameters. Following this, we explain the defined uncertainty scenarios.
\vspace{-1em}
\subsection{System Architecture and Parameters}
\begin{figure}[htbp]
\centering
\includegraphics[width=0.48\textwidth]{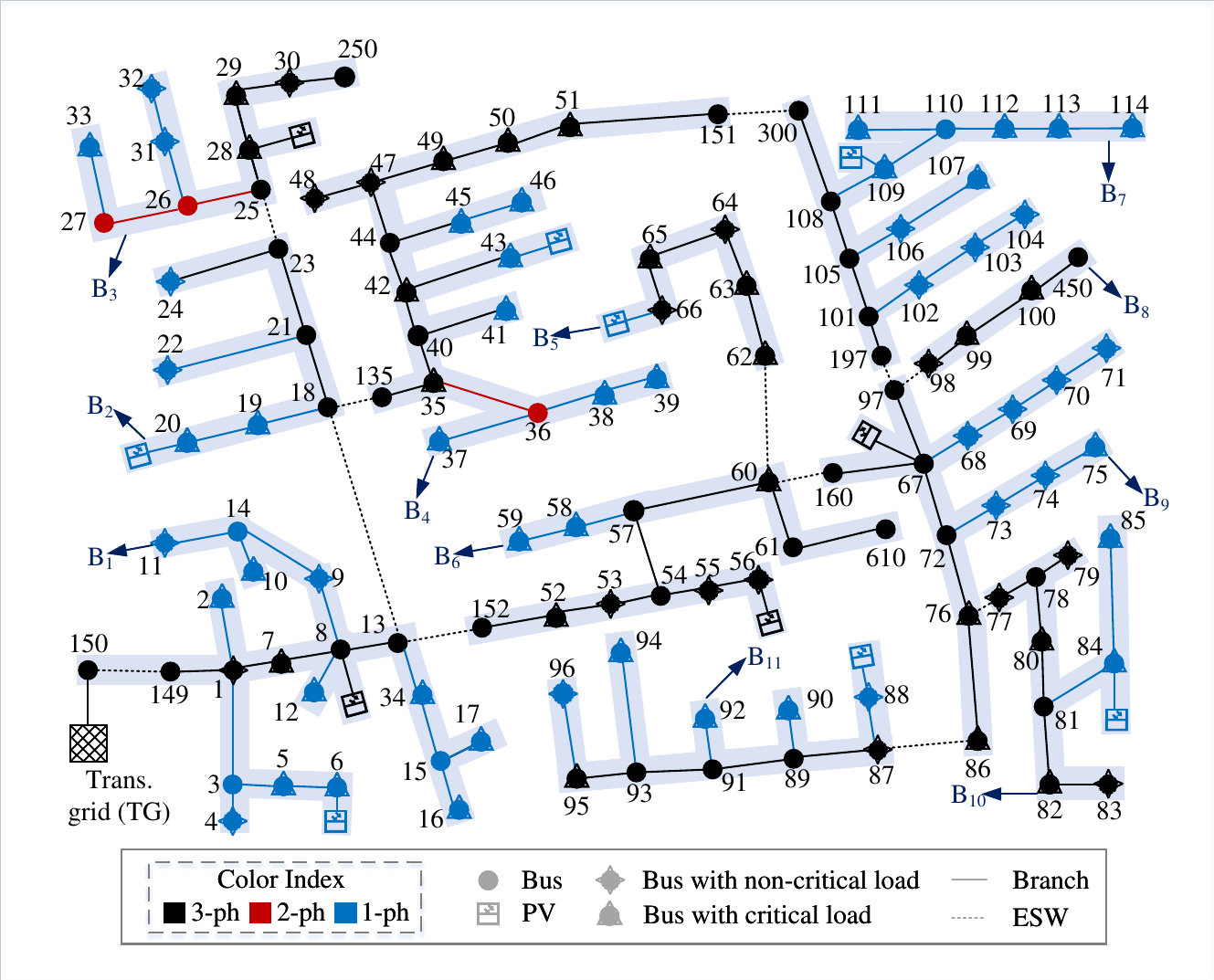}
% \vspace{-1em}
\caption{Modified IEEE 123-node feeder diagram.}
\label{fig:IEEE123diagram}
\end{figure}

The three-phase unbalanced IEEE 123-node feeder is employed to design the BS plan for the DS. The modified DS diagram is shown in Fig.~\ref{fig:IEEE123diagram}. 
The system comprises 88 buses connected to loads, of which 52 are critical loads. The maximum active load demand of the entire system is approximately 3.49 MW, while all loads within the system are assumed to share a uniform power factor $\theta = 0.484$. Behind-the-meter PV systems, with a total installed capacity of 965 kW, account for 28\% of the system's active load demand.

As depicted in Fig.~\ref{fig:IEEE123diagram}, the test system is divided into 12 sections, including 11 segments ($\mathrm{B}_i, i\in\mathcal{G}$) and the TG, separated by 12 ESW, represented by the dashed line. During the resource allocation stage, a three-phase bus within selected segments is chosen as the site for installing GFMI-based BESSs. The maximum allowable rated power and capacity for the BESSs, constrained by the DS budget, are 6.5 MW and 10 MWh, respectively, which also represent the maximum limits for a single BESS.  Conversely, the minimum rated power and capacity for a single BESS, limited by technology constraints, are 0.1 MW and 1 MWh, respectively.

Key parameters for the VSG controller of the BESS are as follows: inertia constant $H=8$ s, PLL time constant $T^{\mathrm{PLL}}=0.05$ s, damping factor $D^p=1$ (per unit), and droop gain $K^{f-p}=89$ (per unit). The safe operation ranges for the frequency, QSS frequency, frequency nadir, and maximum RoCoF, are $(59.50\sim 60.50)$ Hz, $(59.50\sim 60.50)$ Hz, $(57.80\sim 61.80)$ Hz, and $(-4.00\sim 4.00)$ Hz/s, individually.
\vspace{-1em}
\subsection{Generated Scenarios}\label{sse:scenarios}
\begin{figure}[htbp]
\centering
\includegraphics[width=0.48\textwidth]{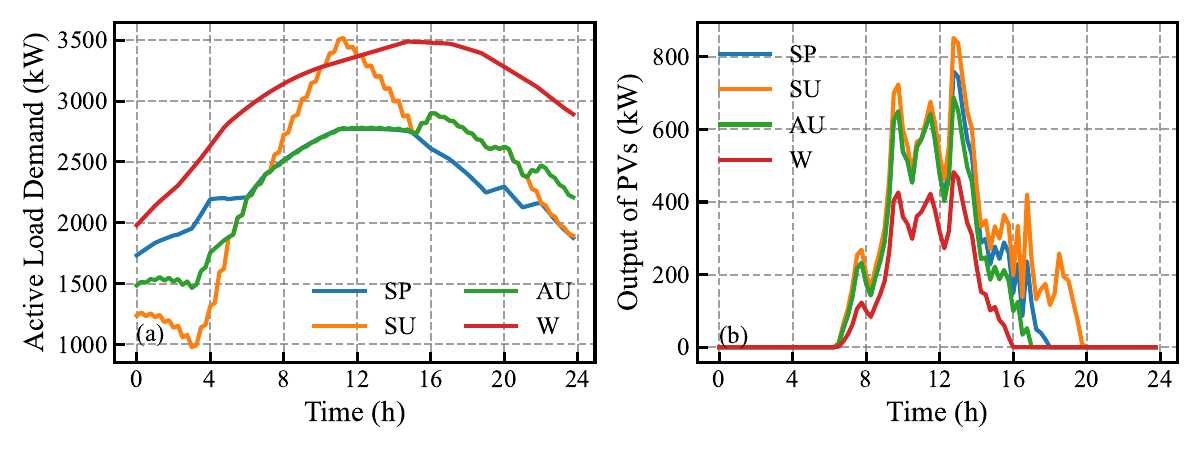}
\vspace{-1em}
\caption{Season-based load and PV curves. (a) Active load demand. (b) PV output. (SP: Spring, SU: Summer, AU: Autumn, W: Winter)}
\label{fig:SeasonLoadPV}
\end{figure}

%The considered uncertainties during the blackstart include the output of PV systems, the load demand, and the outage duration of TG. 
Uncertainties considered during the BS process include PV system output, load demand, and TG outage duration. The PV output and load demand uncertainties follow Beta and Normal distributions, respectively. Season-based characteristic curves for PV output and load demand are derived using sampling and clustering techniques, as illustrated in Fig.~\ref{fig:SeasonLoadPV}.

The TG outage duration uncertainty is modeled using discrete probabilities obtained from historical data, as shown in Table.~\ref{tab:ODP}.
\vspace{-1em}
\begin{table}[htbp]
\centering
\caption{Outage Duration Probability}\label{tab:ODP}
\begin{tabular}{c c c c c}
\toprule
Duration (min) & 60 & 120 & 180 & 240\\
\midrule
Probability & 0.1340 & 0.4290 & 0.2733 & 0.1637\\
\bottomrule
\end{tabular}
\end{table}

Assuming equal probabilities for season-based characteristic curves, the combined probabilities of the 16 scenarios, defined by season $\sigma\in\{\mathrm{SP}, \mathrm{SU}, \mathrm{AU}, \mathrm{W}\}$ and TG outage duration $\nu\in\{60,120,180,240\}$, are given by:
\begin{equation}
    \pi_o = \pi_\sigma\pi_\nu =  0.25\pi_\nu, \forall o = \{\sigma,\nu\}.
\end{equation}
\section{Results}\label{se:5}
In this section, firstly, the stochastic plan is presented and compared with deterministic plans in Case 1. Secondly, the detailed BS process of the stochastic plan under the extreme scenario is illustrated in Case 2. Finally, the frequency security across all scenarios is analyzed in Case 3.
\subsection{Case 1: Risk-averse Allocation Results}
The stochastic SDMG-BS resource allocation model is solved using the 16 scenarios described in Section~\ref{sse:scenarios}, and the risk-averse allocation results are summarized in Table~\ref{tab:AllocationResults}. The stochastic plan allocates three BESSs at bus 18, 62, and 98, with the rated power and capacity of each BESS determined based on trade-offs across all scenarios. Additionally, three ESWs at branches (151, 300), (60, 160), and (150, 149) are replaced with SSWs to enable location-independent synchronizations during the BS process.
\begin{table}[htbp]
\centering
\caption{Risk-averse Resource Allocation Results}\label{tab:AllocationResults}
\begin{tabular}{c c c c}
\toprule
Resource & Location & Power (MW) & Capacity (MWh)\\
\midrule
BESS & Bus 18 & 2.294 & 3.942\\
BESS & Bus 62 & 1.283 & 2.471\\
BESS & Bus 98 & 2.222 & 3.587\\
SSW & Branch (151, 300) & / & / \\
SSW & Branch (60, 160) & / & / \\
SSW & Branch (150, 149) & / & / \\
\bottomrule
\end{tabular}
\end{table}

\begin{figure}[htbp]
\centering
\includegraphics[width=0.48\textwidth]{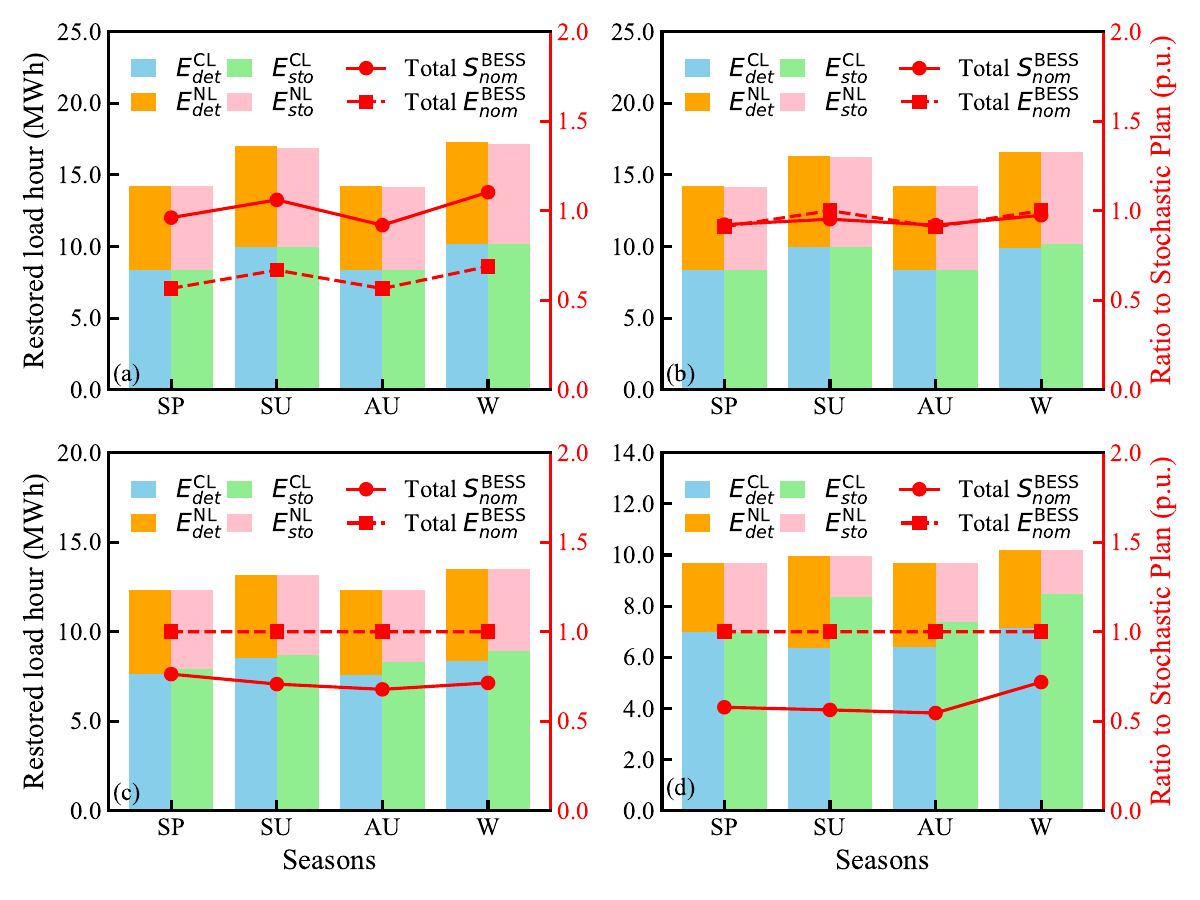}
\vspace{-1em}
\caption{Comparisons between the deterministic plans and stochastic plan. (a) Outage duration is 60 mins. (b) Outage duration is 120 mins. (c) Outage duration is 180 mins. (d) Outage duration is 240 mins.}
\label{fig:Allocationcomparison}
\end{figure}

Furthermore, the restored critical load $E^{\mathrm{CL}}$ and non-critical load $E^{\mathrm{NL}}$ from 8:45 to 13:45 with 15 minutes time-step, obtained using deterministic and stochastic methods across all scenarios, are illustrated in Fig.~\ref{fig:Allocationcomparison}. The left black y-axis shows the restored load, while the right red y-axis depicts the ratios of the total allocated rated power and capacity of the deterministic plans to the stochastic plan.

As shown in Fig.~\ref{fig:Allocationcomparison} (a) and (b), these total allocated rated power ratios greater than one show that the stochastic plan can restore the same amount of non-critical load with a smaller rated power budget compared to deterministic methods. Furthermore, the larger allocated rated power obtained by the stochastic plan also improves the recovery performance of the critical load in the prolonged outage scenarios shown in Fig.~\ref{fig:Allocationcomparison} (c) and (d), where the total allocated rated capacity ratio equal to one indicates that those two methods fully utilize deficient BS capacity budgets.
\subsection{Case 2: Optimal BS Process under Extreme Scenario}
\begin{figure*} [htbp]
    \centering
    \includegraphics[width=\linewidth]{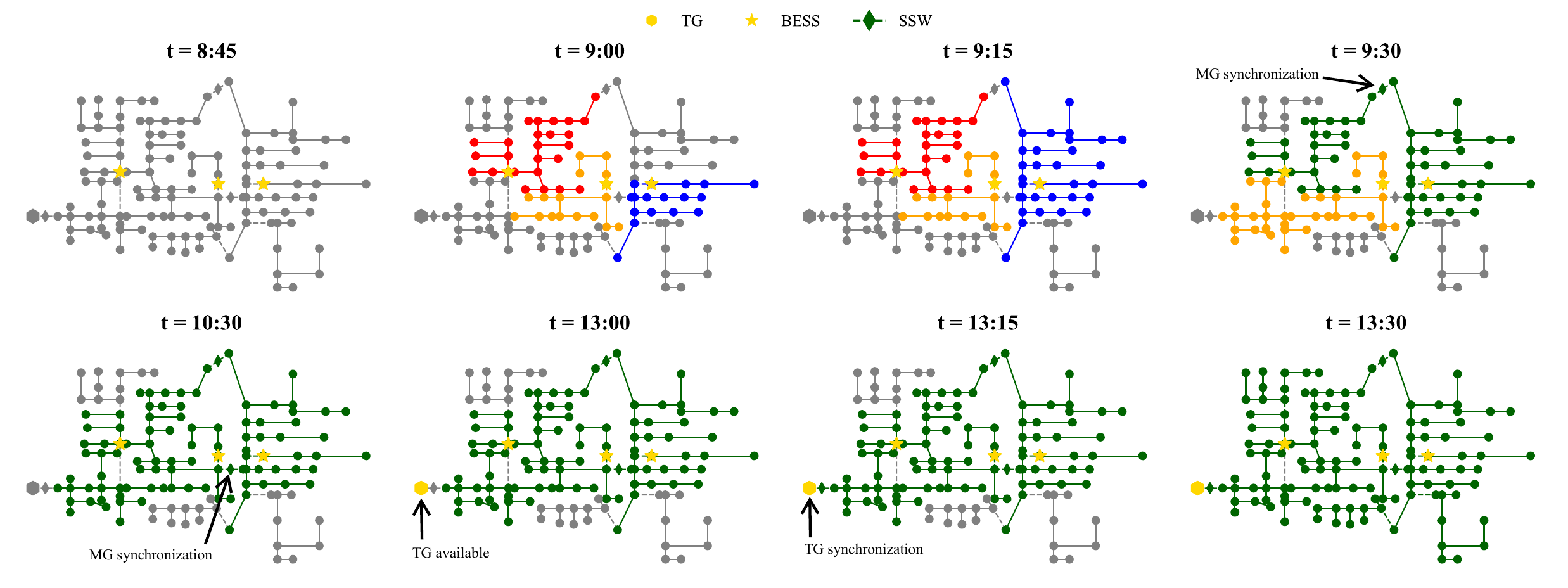}
    \vspace{-2em}
    \caption{Optimal BS cranking path of IEEE 123-node feeder under the extreme scenario.}
    \label{fig:restoreprocessfeeder}
\end{figure*}

The optimal BS process obtained using the stochastic plan for the worst-case scenario (restoring in winter with 240 mins TG outage duration) is shown in Fig.~\ref{fig:restoreprocessfeeder}. Restoration begins at 8:45 and ends at 13:45, with updates at 15-minute intervals.

\begin{figure}[htbp]
\centering
\includegraphics[width=0.48\textwidth]{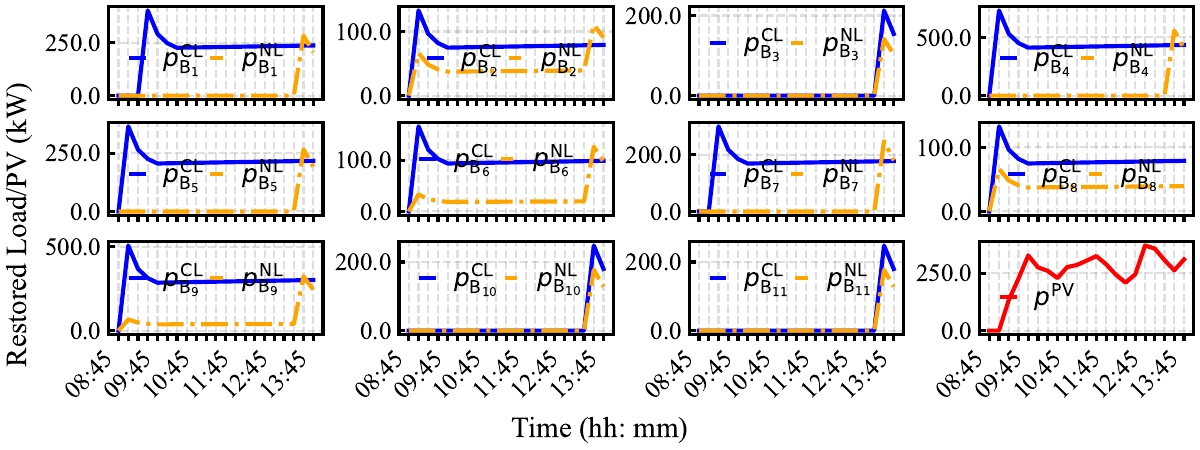}
\vspace{-1em}
\caption{Restored load in each segment and PV output over time.}
\label{fig:restoredseglodandpv}
\end{figure}

As depicted in Fig.~\ref{fig:restoreprocessfeeder}, without employing the SDMG-BS framework, the DS would have remained without power until 13:00 due to the loss of TG. However, as shown in Fig.~\ref{fig:restoreprocessfeeder}, with GFMI capability, BS was initiated at 8:45 by activating GFMIs-based BESSs at bus 18, 62, and 98, represented with gold stars. These BESSs served as the starting points for cranking paths to energize other GFLIs-based behind-meter PVs. By 9:00, the segments containing the BESSs, $\mathrm{B}_2$, $\mathrm{B}_5$, and $\mathrm{B}_8$, and their corresponding nearby segments, $\mathrm{B}_4$, $\mathrm{B}_6$, and $\mathrm{B}_9$, were energized to separately form three isolated MGs, colored as red, orange, and blue, respectively, through closing ESWs (18, 135), (60, 62), and (97, 98). Then, the MG dominated by the BESS at bus 98 expanded its boundary by energizing the $\mathrm{B}_7$ through closing ESW (97, 197) until 9:15. At 9:30, the two MGs, ruled by the BESSs at bus 18 and 98 and colored as green, were synchronized by closing the SSW (151, 300), increasing the system's energy and power capacity to restore more PVs and loads. Meanwhile, the $\mathrm{B}_1$ was restored by the MG governed by the BESS at bus 62 through acting the ESW (13, 152). One hour later, this MG was synchronized with those two merged MGs by closing the SSW (60, 160), where the GFMI and GFLI resources were shared to maintain the restored segments. This configuration remained in place until 13:00 when the TG returned online and turned its color from grey into gold. However, it wasn't connected until 13:15. At that point, the restored island MG was synchronized with the TG, which picked up the outage $\mathrm{B}_3$, $\mathrm{B}_{10}$, and $\mathrm{B}_{11}$ at 13:30 and realized the final configuration capable of operating continuously moving forward. The detailed critical and non-critical load restoration profiles segment by segment and the total restored PV output are shown in Fig.~\ref{fig:restoredseglodandpv}.

\begin{figure}[htbp]
\centering
\includegraphics[width=0.48\textwidth]{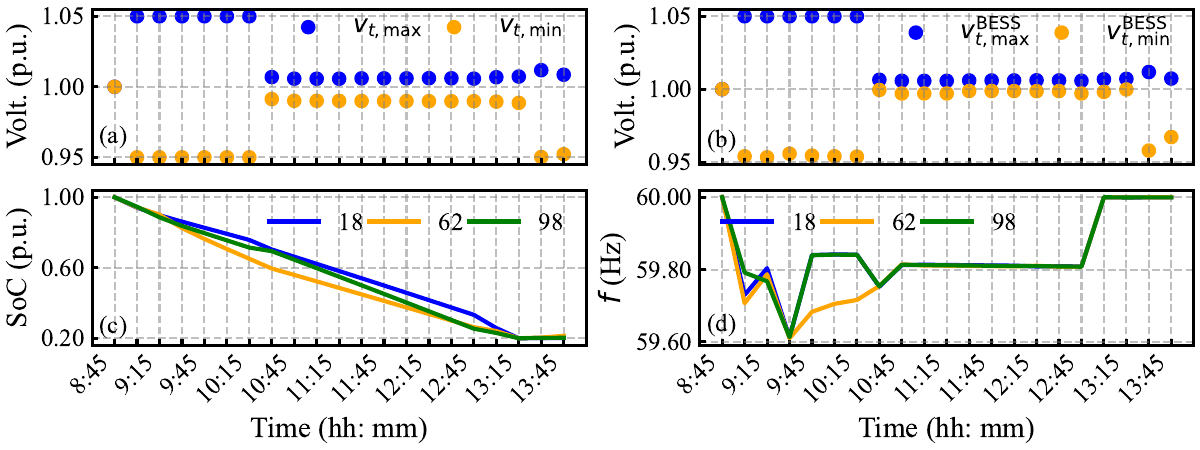}
\vspace{-1em}
\caption{System performance. (a) Sytem nodal voltage level. (b) BESS nodal voltage level. (c) SoCs of the BESSs. (d) Frequencies of the BESSs.}
\label{fig:voltageandfrequency}
\end{figure}

\vspace{-0.2cm}
The system performances during the BS are displayed in Fig.~\ref{fig:voltageandfrequency}.

The maximum and minimum nodal voltages at each time step across all buses and BESSs in Fig.~\ref{fig:voltageandfrequency} (a) and (b) are restricted into the safe operation range $(0.95\sim 1.05)$ in per unit as required in Eq.~\ref{eq:powerflowsecurity3}. Similarly, the SoC and frequency of each BESS during the BS are also limited as Eq.~\ref{eq:SoCconstraint} and Eq.~\ref{eq:frequencysecurityfrequency} asked. Moreover, the synchronization moments between restored MGs and with the TG can be captured with the intersection points in Fig.~\ref{fig:voltageandfrequency} (d).
\subsection{Case 3: Frequency Security Analysis}
\begin{figure}[htbp]
\centering
\includegraphics[width=0.48\textwidth]{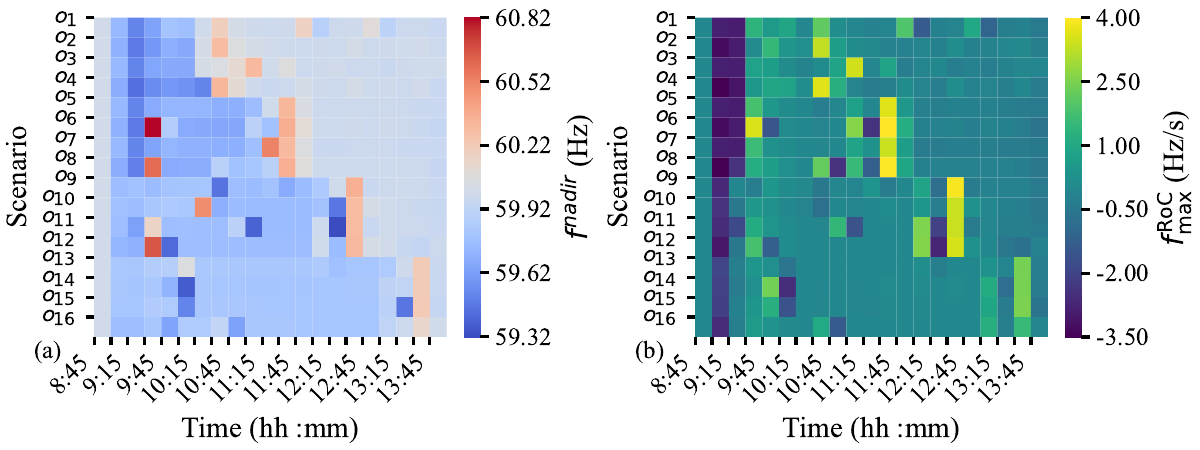}
\vspace{-1em}
\caption{Frequency security across all scenarios. (a) Frequency nadir. (b) Maximum RoCoF.}
\label{fig:frequencysecurity}
\end{figure}

The frequency security during the BS is analyzed and presented in Fig.~\ref{fig:frequencysecurity} by collecting the transient frequency indices obtained by the two-stage stochastic model across all deterministic scenarios.

Based on the relationships among the operation frequency, QSS frequency, and frequency nadir in Eq.~\ref{eq:frequencyindices}, the investigations on the frequency nadir and maximum RoCoF are enough to reveal the transient frequency performance of the GFMI-based BESS during the BS. As shown in Fig.~\ref{fig:frequencysecurity} (a), the frequency nadir obtained by the stochastic allocation plan during the BS over all deterministic scenarios are located within $(59.32\sim 60.82)$ Hz, which is included in the safe operation range $(57.80\sim 61.80)$ Hz. Furthermore, the maximum RoCoFs of each BESS allocated by the stochastic plan during the BS over all deterministic scenarios have the level between -3.50 and 4.00 Hz/s, which is also included in the safe range $(-4.00\sim 4.00)$ Hz/s.
\section{Conclusion}\label{se:6}
This paper presents a two-stage stochastic resource allocation model with frequency constraints for the SDMG-BS framework, designed to facilitate the restoration of DER-dominated DSs after prolonged outages. The model incorporates multi-source uncertainties, including season-dependent operational statuses of RES and loads and varying TG outage durations, to optimize resource allocation. Additionally, the SDMG-BS framework enables flexible recovery strategies through location-independent synchronization between restored MGs and the TG, facilitated by the operation of SSWs. The integration of VSG-controlled GFMIs ensures frequency security throughout the BS process. The key findings are summarized as follows. First, the stochastic allocation plan enables the resilient and efficient restoration of the BS strategy under diverse operational conditions, optimizing resource utilization. Second, restoration can be initiated from multiple locations by forming islanded MGs, which support each other through flexible synchronization enabled by the SDMG-BS framework. Moreover, Frequency stability in the DS is maintained by constraining the transient frequency indices of GFMI-based BESSs within safe ranges across all scenarios. Future work will investigate the frequency dynamics of rotating loads and the impact of protection systems on the BS strategy to further refine and enhance the proposed approach.
% or


%\appendices
\appendix[Voltage Controller Deviation of VSG]\label{ap:VSG}
% \section{}\label{ap:VSG}
Terminal voltage of GFMI deviating from the base value ($V^b$) with reactive power output ($q$) and voltage droop gain ($K^{V-q}$) is expressed as follows:
\begin{equation}\label{eq:voltageVSG}
    V= V^b - K^{V-q} q
\end{equation}
Given that $v=V^2$ and substituting $V$ intoEq.~\ref{eq:voltageVSG}, we get:
\begin{align}
    v &= (V^b-K^{V-q} q)^2\notag\\
      &= (V^b)^2 - 2V^b K^{V-q} q + (K^{V-q} q)^2\notag\\
      &= (V^b)^2 + \Delta v^{inc}.
\end{align}
Here, a new variable $v^{inc}$ representing two non-linear terms is introduced for brevity.

%\section*{Acknowledgment}
\ifCLASSOPTIONcaptionsoff
  \newpage
\fi
\bibliographystyle{IEEEtran}
\bibliography{IEEEabrv, references.bib}
\end{document}